\documentclass[aps,prd,12pt,nofootinbib,notitlepage]{revtex4-1}

\usepackage{amsfonts}
\usepackage{amsmath,epsfig,bm}
\usepackage{graphicx}
\usepackage{bm}
\usepackage{color}
\renewcommand{\vec}[1]{\mbox{\boldmath$#1$}}

\usepackage{tabularx} 

\usepackage{times}

\newcommand{\be}{\begin{equation}}
\newcommand{\ee}{\end{equation}}
\newcommand{\ba}{\begin{eqnarray}}
\newcommand{\ea}{\end{eqnarray}}
\newcommand{\bd}{\begin{displaymath}}
\newcommand{\ed}{\end{displaymath}}

\def\thalf{{\textstyle{\frac{1}{2}}}}

\def\oneqt{{\textstyle{\frac{1}{4}}}}

\def\rt3{\sqrt{3}}
\def\rt6{\sqrt{6}}
\def\mL2{(m_{\phi}L)^2}

\begin{document}

\title{{\bf Spin versus Helicity Equilibration Times and Lagrangian for Strange Quarks in Rotating Quark-Gluon Plasma}}
\author{Joseph I. Kapusta$^1$, Ermal Rrapaj$^{1,2}$, and Serge Rudaz$^1$}
\affiliation{$^1$School of Physics and Astronomy, University of Minnesota, Minneapolis, Minnesota 55455, USA \\
$^2$Department of Physics, University of California, Berkeley, CA 94720, USA}

\vspace{.3cm}

\parindent=20pt

\begin{abstract}
Measurements of the net polarization of $\Lambda$ and $\bar{\Lambda}$ hyperons at the Relativistic Heavy Ion Collider (RHIC) have stimulated much interest in how strange quarks might align their spin with the vorticity of the matter created in heavy ion collisions.  We calculate the Lagrangian in the rest frame of a fluid element undergoing rotation with angular velocity $\omega$ including photon and gluon fields.  There is an additional coupling between the quarks and the gauge fields proportional to $\omega$, but this vertex does not change the spin of the quarks.  We also show that the times to equilibrate quark helicity and spin parallel to the vorticity are the same so long as $\omega$ is small compared to the temperature.
 
\end{abstract}
\date{\today}

\maketitle

\section{Introduction}

Polarization of $\Lambda$ and $\bar{\Lambda}$ hyperons was proposed as an observable that provides information on the vorticity of the hot, dense matter created in non-central heavy ion collisions \cite{Wang1,Becattini1}.  In these collisions the spins of the $\Lambda$ and $\bar{\Lambda}$ ought to couple to the vorticity, resulting in a splitting in energy between particles with spin parallel and antiparallel to the vorticity.  The distribution of their decay products can be used to infer their polarizations.  Measurements of these polarizations have been made by the STAR collaboration over the full range of beam energies at RHIC \cite{FirstSTAR,Nature,SecondSTAR}.  These measurements indicate that $\omega = (9 \pm 1) \times 10^{21}$ s$^{-1}$, with a systematic error of a factor of two, when averaging over the entire RHIC energy range.   This converts to an energy of $\omega = 6 \pm 1$ MeV.  It was concluded that RHIC produces matter with the highest vorticities ever observed.    

Similar to the still unsettled question of how quarks and gluons come to thermal equilibrium in heavy ion collisions is the dynamical mechanism by which the hyperons become polarized.  In the quark model the spin of the $\Lambda$ is carried by the strange quark \cite{Jennings1,Cohen}.  One possibility is that the strange quarks become polarized in the quark-gluon plasma phase and pass that poalrization on to the $\Lambda$ hyperons during hadronization.  If that is the case, how did the strange quarks become polarized to begin with?  Were they created with a polarization which did not change much until hadronization?  Or were they created unpolarized and only acquired it during the subsequent evolution of the quark-gluon plasma?  The answers to these questions rely on the magnitude of the spin equilibration time.  If the equilibration time is long then the strange quarks carry memory of their polarization at the time of their creation.  If the equilibration time is short then they represent the conditions of the system at the time of hadronization.  

In Ref. \cite{KRR1} we considered two mechanisms by which strange quark spin could equilibrate.  The first mechanism recognizes that there will be fluctuations in the direction and magnitude of the vorticity in heavy ion collisions.  These fluctuations will drive the spins back towards equilibrium, just as fluctuations around a constant magnetic field drive electron spins towards equilibrium.  The second mechanism considers the scattering of massive strange quarks with massless up and down quarks and gluons in the plasma to lowest order in perturbation theory.  Since helicity is conserved in QCD interactions when the quark is massless, helicity flip can only occur when the quark has a mass.  Both mechanisms resulted in equilibration times far too long to be relevant to heavy ion collisions.  In Ref. \cite{KRR2} we considered the Nambu--Jona-Lasinio model with the inclusion of the six-quark Kobayashi--Maskawa--'t Hooft interaction which breaks axial U(1) symmetry.  Using instanton inspired models for the temperature dependence of the axial symmetry breaking, we found that constituent strange quarks can reach spin equilibrium at temperatures below about 170 MeV, just before they hadronize to form hyperons.

In this paper we address three questions left unanswered in our previous works \cite{KRR1,KRR2}.  Is there an additional coupling between quarks and gauge fields that would flip the spin or helicity?  The answer is no.  Are the equilibration times for helicity and spin parallel to the vorticity the same?  The answer is yes, so long as the vorticity is small compared to the temperature, which is the case in heavy ion collisions.  Is the equilibration time affected by vorticity?  The answer is yes, but insignificant in heavy ion collisions. 

\section{Tetrads and Lagrangians}

Consider a fluid element undergoing rotation.  See our previous work for details \cite{KRR1}.  Here we only recall the essential formulae needed to address the questions posed in this paper.  

The idea is to set up an inertial coordinate system at rest with respect to a fluid element at every space-time point.  Let $x^{\mu}$ represent the space-time coordinates of an observer at rest in the fluid element and $\xi^a$ the coordinates of an inertial frame.  Then
\be
 g_{\mu\nu}(x) dx^{\mu} dx^{\nu} = \eta_{ab} d\xi^a d\xi^b  \,.
\ee
When there is no cause for confusion we use Greek indices for the $x$-coordinates, Latin indices $a, b, ...$ for the $\xi$-coordinates, and Latin indices $i, j, ...$ for spatial indices.  The Minkowski metric is $\eta_{ab} = {\rm diag}(1,-1,-1,-1)$.  The tetrad is defined as
\be
e_{\mu}^{\;\;a}(x) = \frac{\partial \xi^a}{\partial x^{\mu}}
\ee
while the inverse tetrad is 
\be
e^{\mu}_{\;\;a}(x) = g^{\mu\nu}(x) \eta_{ab} e_{\nu}^{\;\;b}(x) \,.
\ee
The tetrads obey the orthogonality properties
\ba
e_{\mu}^{\;\;a}(x) e^{\mu}_{\;\;b}(x) &=& \delta^a_b \nonumber \\
e^{\mu}_{\;\;a}(x) e_{\nu}^{\;\;a}(x) &=& \delta^{\mu}_{\nu} \,.
\ea
The Dirac matrices $\hat{\gamma}^{\mu}(x)$ become space-time dependent.  They are obtained from the usual Dirac matrices $\gamma^a$ by
\be
\hat{\gamma}^{\mu}(x) = e^{\mu}_{\;\;a}(x) \gamma^a \,.
\ee
They satisfy
\be
\hat{\gamma}^{\mu} \hat{\gamma}^{\nu} + \hat{\gamma}^{\nu} \hat{\gamma}^{\mu} = 2 g^{\mu\nu}
\ee
compared to
\be
\gamma^{a} \gamma^{b} + \gamma^{b} \gamma^{a} = 2 \eta^{ab} \,.
\ee
One finds that the gradient of a spinor is replaced by a covariant derivative.
\be
\partial_{\mu} \psi \rightarrow D_{\mu} \psi = \left(\partial_{\mu} + \Gamma_{\mu} + i e A_{\mu}\right) \psi
\ee
Here $A_{\mu}$ is the electromagnetic vector potential.  The symbol $\Gamma_{\mu}$ is called the spin connection.  The Dirac equation is \cite{Brill}
\be
\left[ i \hat{\gamma}^{\mu}(x) D_{\mu} - m \right]\psi = 0 \,.
\ee

Consider a region of space where a fluid element is rotating in an anti-clockwise sense around the $z$ axis with angular speed $\omega$ which may be considered constant within that region.  We follow Ref. \cite{Hehl} and choose the tetrad as the $4\times 4$ matrix
\be
e_{\mu}^{\;\;a}(x) =
\begin{pmatrix}
1 & v_x & v_y & 0 \\
0 & 1 & 0 & 0 \\
0 & 0 & 1 & 0 \\
0 & 0 & 0 & 1 \\
\end{pmatrix}
\ee
where $v_x = -\omega y$, $v_y = \omega x$, and $v_z = 0$.  From this is it straightforward to find the metric
\be
g_{\mu\nu}(x) =
\begin{pmatrix}
1 -v^2 & -v_x & -v_y & 0 \\
-v_x & -1 & 0 & 0 \\
-v_y & 0 & -1 & 0 \\
0 & 0 & 0 & -1 \\
\end{pmatrix} \, .
\ee
The nonzero components of the affine connection are
\ba
\Gamma^1_{00} &=& \omega^2 x \nonumber \\
\Gamma^2_{00} &=& \omega^2 y \nonumber \\
\Gamma^2_{01} &=& \omega \nonumber \\
\Gamma^1_{02} &=& -\omega \,.
\ea
The only nonzero component of $\Gamma_{\mu}$ is
\be
\Gamma_0 = -\frac{i}{2} \omega \Sigma_3 = -\frac{i}{2} \omega 
\begin{pmatrix}
\sigma_3 & 0 \\
0 & \sigma_3 \\
\end{pmatrix} \,.
\ee 
Finally the Dirac matrices are $\hat{\gamma}^{\mu}(x) = \gamma^{\mu} - v^{\mu}(x) \gamma^0$.

The single particle Hamiltonian can be found by writing the Dirac equation in the form $i \partial_0 \psi = H \psi$ with the result
\be
H = \beta m +eA^0 +\alpha^j ( -i \partial_j - eA^j) - \omega [ x (-i \partial_2 -eA^2) - y (-i \partial_1-eA^1)] - \thalf \omega \Sigma_3 \,.
\label{spH}
\ee
Defining the vorticity
\be
\thalf \nabla \times \vec{v} = \vec{\omega}
\ee
we can express the Hamiltonian in terms of the orbital and spin angular momentum as
\be
H = \beta m + eA^0 + \vec{\alpha} \cdot ( {\bf p} - e{\bf A}) - \vec{\omega} \cdot [{\bf x} \times ( {\bf p} - e{\bf A}) + {\bf S}] 
\label{HLS}
\ee
where ${\bf p} = -i \nabla$.  It can also be written as
\be
H = \beta m + eA^0 + (\vec{\alpha} - {\bf v}) \cdot ( {\bf p} - e{\bf A}) - \vec{\omega} \cdot {\bf S} \,.
\label{HpS}
\ee
When taking the nonrelativistic limit via the Foldy-Wouthuysen procedure, it is known that the orbital angular momentum term gives rise to the usual Coriolis and centrifugal forces \cite{Obukhov2013,Matsuo2017}.  The last term is the spin-rotation coupling.

The conserved current density is
\be
j^{\mu} = \overline{\psi} \hat{\gamma}^{\mu} \psi \,.
\ee
One finds by direct calculation from the Dirac equation that
\be
\partial_{\mu} j^{\mu} = 0 \,.
\ee
The piece of the Lagrangian that leads to the Dirac equation can be written in the local rest frame as
\be
{\cal L} = \overline{\psi} \left[ i \gamma^{\mu} ( \partial_{\mu} + i e A_{\mu} ) - m -
i \gamma^0 v^{\mu} ( \partial_{\mu} + i e A_{\mu} ) + \oneqt \gamma^0 \epsilon_{0\alpha\beta\kappa} \sigma^{\alpha\beta} \omega^{\kappa} \right]
\psi
\label{LA}
\ee
which uses the Minkowski metric and $v^{\mu} = (0, v_x, v_y, 0)$.  To include a chemical potential $\mu$ one adds the term $\overline{\psi} ( - \mu \gamma^0 ) \psi$.  To include SU(N) gauge fields the derivative $\partial_{\mu} + i e A_{\mu}$ appearing in Eq. (\ref{LA}) should be changed to $\partial_{\mu} + i e A_{\mu} + i g A^a_{\mu} G^a$ where $G^a$ are the generators of the group.  It is apparent from both the Dirac equation and the Lagrangian that there is an additional coupling between the fermion field and the gauge field proportional to $\omega$, but it does not involve the tensor $\sigma^{\alpha\beta}$.  This is in contrast to the phenomenological coupling proposed in Ref. \cite{Ayala} for QCD.

Now we consider what happens to the electromagnetic field in this formalism.  See also Ref. \cite{Arendt}.  The electric and magnetic fields are obtained from the field strength tensor as $F^{30} = E_z$, where $E_z$ is the $z$ component of the ordinary electric field vector ${\bf E}$, $F^{12} = - B_z$ for the magnetic field, and similarly for the other components.  Maxwell's equations are
\bd
\partial_{\alpha} F^{\alpha \beta} = j^{\beta}
\ed
\be
\partial_{\alpha} F_{\beta \gamma} + \partial_{\beta} F_{\gamma \alpha} + \partial_{\gamma} F_{\alpha \beta} = 0
\ee
These are true in either frame of reference because in both cases $g= \det(g_{\mu\nu}) = -1$.

Let $\bar{F}^{ab}$ denote the field strength tensor in the inertial frame and $F^{\mu\nu}$ denote it in the local frame.  Using $F^{\mu\nu} = e^{\mu}_a e^{\nu}_b \bar{F}^{ab}$ we find that 
\be
F^{\mu\nu}=\bar{F}^{\mu\nu}+v^{\mu}\bar{F}^{\nu 0}-v^{\nu}\bar{F}^{\mu 0}.
\ee
The individual components are $F^{i0} = \bar{F}^{i0}$ and $F^{ij} = \bar{F}^{ij} +v^{i}\bar{F}^{j0}-v^{j}\bar{F}^{i 0}$ or, equivalently, ${\bf E} = \bar{{\bf E}}$ and ${\bf B} = \bar{{\bf B}} - {\bf v} \times \bar{{\bf E}}$.  The gauge field contribution to the Lagrangian is unchanged: $-\oneqt F_{\mu\nu} F^{\mu\nu} = -\oneqt \bar{F}_{ab} \bar{F}^{ab}$.  In addition are the usual gauge fixing terms.  It should be obvious that this can be generalized to SU(N) gauge fields.  The gauge field kinetic energy has the usual form $-\oneqt F_{\mu\nu}^a F^{\mu\nu}_a$ plus gauge fixing terms.  The only difference is that the components of the field strength tensor are not gauge invariant, unlike in the Abelian theory. 

Finally we verify the expected relationship between the vector potentials.  Starting with the expression in the inertial frame we have
\be
\bar{F}^{ab} = \partial^a \bar{A}^b - \partial^b \bar{A}^a 
= \left( \frac{\partial}{\partial x_{\alpha}} \bar{A}^b \right) \frac{\partial x_{\alpha}}{\partial \xi_a}
- \left( \frac{\partial}{\partial x_{\beta}} \bar{A}^a \right) \frac{\partial x_{\beta}}{\partial \xi_b}
= e_{\alpha}^{\;\;a} \, \partial^{\alpha} \bar{A}^b - e_{\beta}^{\;\;b} \, \partial^{\beta} \bar{A}^a \,.
\ee
Multiply this expression by $e^{\mu}_{\;\;a} e^{\nu}_{\;\;b}$ and use the identity $e^{\mu}_{\;\;a} e_{\alpha}^{\;\;a} = \delta^{\mu}_{\alpha}$ to get
\ba
F^{\mu\nu} &=& e^{\mu}_{\;\;a} e^{\nu}_{\;\;b} \bar{F}^{ab} = e^{\nu}_{\;\;a} \, \partial^{\mu} \bar{A}^a
- e^{\mu}_{\;\;a} \, \partial^{\nu} \bar{A}^a \nonumber \\
&=& \partial^{\mu} \left( e^{\nu}_{\;\;a} \bar{A}^a\right) - \partial^{\nu} \left( e^{\mu}_{\;\;a} \bar{A}^a\right)
- \bar{A}^a \left( \partial^{\mu} e^{\nu}_{\;\;a} - \partial^{\nu} e^{\mu}_{\;\;a} \right) \,.
\ea
The last term vanishes by symmetry because $e^{\mu}_{\;\;a} = \partial^{\mu} \xi_a$.  Hence $A^{\mu} = e^{\mu}_{\;\;a} \bar{A}^a$ as expected.  It is obvious that this generalizes to an SU(N) gauge field.

\section{Relation between Helicity Flip and Spin Flip}

A common approximation to the Boltzmann equation is the energy-dependent relaxation time approximation.  Consider the reaction $a+b \rightarrow c+d$.  Let us suppose that all species of particles for all values of momentum are in equilibrium except for species $a$ with momentum ${\bf p}_a$.  Replace all phase space distributions $f$ with their equilibrium values $f^{\rm eq}$ except for $f_a$, which we allow to be out of equilibrium by a small amount.  Thus we write $f_a = f_a^{\rm eq} + \delta f_a$ and  
\be
\frac{\partial f_a({\bf x},t,{\bf p}_a)}{\partial t} + {\bf v}_a \cdot \nabla f_a({\bf x},t,{\bf p}_a) =
- \frac{1}{\tau_a(E_a)} \delta f_a({\bf x},t,{\bf p}_a) \,.
\label{relaxBoltzeq}
\ee
The equilibration time is determined by \cite{ChakrabortyKapusta2011,AlbrightKapusta2016}
\be
\frac{1+d_a f_a^{\rm eq}}{\tau_a(E_a)}  =  \sum_{bcd} \frac{{\cal N}}{1+\delta_{ab}}
\int \frac{d^3p_b}{(2\pi)^3} \frac{d^3p_c}{(2\pi)^3} \frac{d^3p_d}{(2\pi)^3} \,  W(a,b|c,d) f_b^{\rm eq}  \left( 1 + d_c f_c^{\rm eq} \right) \left( 1 + d_d f_d^{\rm eq} \right)  \,.
\label{eq:RTA:relaxationtime}
\ee
Here $d_i = (-1)^{2s_i}$ corresponding to Bose enhancement or Pauli suppression.  The $W$ is related to the dimensionless amplitude ${\cal M}$ by
\be
W(a,b|c,d) = \frac{(2\pi)^4 \delta^4\left( p_a+p_b-p_c-p_d\right)}
{2E_a2E_b2E_c2E_d} |{\cal M}(a,b|c,d)|^2 \, .
\ee
The $|{\cal M}(a,b|c,d)|^2$ is averaged over spin in both the initial and final states; this compensates the spin factor $2s_i+1$ in the phase space integration.
Finally, the ${\cal N}$ is a degeneracy factor for spin, color, and any other internal degrees of freedom.  Its value depends on how these variables are summed or averaged over in $|{\cal M}|^2$.  

In past papers we found it much easier and expedient to compute the helicity flip equilibration time in the absence of vorticity than to compute the time for equilibration of the spin component parallel to the vorticity with nonzero vorticity.   The physical argument we gave is that the magnitude of helicity in high energy heavy ion collisions is much less than the temperature, $\omega \ll T$, with $\omega = 6 \pm 1$ MeV and $T \ge 150$ MeV.  We elucidate that point in this section.

First, consider the 2 component Pauli spinor and neglect the small components of the Dirac spinor.  The eigenvector of the helicity operator 
$\vec{\sigma} \cdot \hat{{\bf p}}$ for positive helicity $\lambda = +1$ is
\be
\chi_+ =
\begin{pmatrix}
\cos \left(\frac{\theta}{2}\right) \\
\\
e^{i \phi} \sin \left(\frac{\theta}{2}\right) \\
\end{pmatrix}
\ee 
and for negative helicity $\lambda = -1$ is
\be
\chi_- =
\begin{pmatrix}
- \sin \left(\frac{\theta}{2}\right) \\
\\
e^{i \phi} \cos \left(\frac{\theta}{2} \right)\\
\end{pmatrix}
\ee 
where the angles refer to the direction of the momentum relative to the $z$ axis in polar coordinates.  From these one easily reads off the relationships among the distribution functions.  With an up arrow denoting spin quantized along the $+z$ axis and a down arrow denoting spin quantized along the $-z$ axis they are
\ba
f({\bf p}, +) &=& \cos^2\left(\frac{\theta}{2}\right) f({\bf p}, \uparrow) + \sin^2\left(\frac{\theta}{2}\right) f({\bf p}, \downarrow) \nonumber \\
f({\bf p}, -) &=& \cos^2\left(\frac{\theta}{2}\right) f({\bf p}, \downarrow) + \sin^2\left(\frac{\theta}{2}\right) f({\bf p}, \uparrow)  \,.
\label{f's}
\ea
These relations are intuitively correct.  For example, consider what happens when $\theta$ goes from 0 to $\pi/2$ to $\pi$.  For changes in helicity or spin the particle number is conserved so that
\be
\delta f({\bf p}, +) + \delta f({\bf p}, -) = 0 =
\delta f({\bf p}, \uparrow) + \delta f({\bf p}, \downarrow) \,.
\ee
Thus $\delta f({\bf p}, +) = \cos \theta \, \delta f({\bf p}, \uparrow)$.  Examination of Eq. (\ref{relaxBoltzeq}) then shows that the equilibration times for helicity and for spin quantized along the direction of vorticity are equal.

Next, consider the 4 component Dirac spinor.  The spin operator does not commute with $H$ but the helicity operator does.  Therefore we compare helicity eigenstates to Dirac spinors which have spin quantized in the $z$ direction in the quark's rest frame.  (Note that in experiments the $z$ component of spin is measured in the hyperon rest frame \cite{SecondSTAR}.)  Positive energy spinors with spin parallel or antiparallel to the $z$ axis, as defined by the vorticity and as measured in the quark's rest frame, are
\ba
u_{\uparrow}({\bf p}) = \sqrt{\frac{E+m}{2m}}
\begin{pmatrix}
1  \\
0  \\
\dfrac{p \cos \theta}{E+m}  \\
\dfrac{p \sin \theta \, e^{i\phi}}{E+m} \\
\end{pmatrix}
\:\:\:\:\:\:\:\:\:\:
u_{\downarrow}({\bf p}) = \sqrt{\frac{E+m}{2m}}
\begin{pmatrix}
0 \\
1 \\
\dfrac{p \sin \theta \, e^{-i\phi}}{E+m}  \\
- \dfrac{p \cos \theta}{E+m} \\
\end{pmatrix} \,.
\label{w}
\ea
Positive energy spinors with helicity parallel or antiparallel to the momentum of the quark are
\ba
u_+({\bf p}) = \sqrt{\frac{E+m}{2m}}
\begin{pmatrix}
\cos(\frac{\theta}{2}) \\
\sin(\frac{\theta}{2}) \, e^{i\phi} \\ 
\dfrac{p \cos (\frac{\theta}{2})}{E+m} \\
\dfrac{p \sin(\frac{\theta}{2}) \, e^{i\phi}}{E+m}  \\ 
\end{pmatrix}
\:\:\:\:\:\:\:\:\:\:
u_-({\bf p}) = \sqrt{\frac{E+m}{2m}}
\begin{pmatrix}
- \sin(\frac{\theta}{2})  \\
\cos(\frac{\theta}{2}) \, e^{i\phi} \\
\dfrac{p \sin (\frac{\theta}{2})}{E+m} \\
- \dfrac{p \cos (\frac{\theta}{2})}{E+m} \, e^{i\phi} \\
\end{pmatrix} \,.
\label{u}
\ea
It is easy to check that
\ba
u_+({\bf p}) &=& \cos\left(\frac{\theta}{2}\right) u_{\uparrow}({\bf p}) + \sin\left(\frac{\theta}{2}\right) \, e^{i\phi} u_{\downarrow}({\bf p}) \nonumber \\
u_-({\bf p}) &=& -\sin\left(\frac{\theta}{2}\right) u_{\uparrow}({\bf p}) + \cos\left(\frac{\theta}{2}\right) \, e^{i\phi} u_{\downarrow}({\bf p}) \,.
\ea
Hence the relations in Eq. (\ref{f's}) follow again.  Similar relations among the negative energy states follow in the same way.

\section{Effect of Vorticity on the Equilibrium Distribution Functions and Equilibration Times}

In this paper we are interested in the spin-rotation coupling.  The vorticity couples to the total angular momentum ${\bf J} = {\bf L} + {\bf S}$, and it is ${\bf J}$ which commutes with the Hamiltonian of Eq. (\ref{HLS}) or (\ref{HpS}).  Nevertheless in this section we shall  drop the term $\vec{\omega} \cdot {\bf L} = \vec{v} \cdot {\bf p}$.  Because the vorticity in energy units is so small in high energy heavy ion collisions this appears justifiable.  Alternatively, one may restrict attention to the region near the origin where the orbital angular momentum is small and $|\vec{v}| \ll 1$.  Keeping the coupling of vorticity to orbital angular momentum complicates the problem significantly, and one should perhaps use an angular momentum basis rather than a momentum basis.

Consider the Hamiltonian $H = m \beta + \vec{\alpha} \cdot {\bf p} - \thalf \omega \Sigma_3$.  Define $E = \sqrt{p^2 + m^2}$ and $E_3 = \sqrt{p_3^2 + m^2}$.  The two positive energy states have eigenvalues $E_{\pm} = \sqrt{ E^2 + \oneqt \omega^2 \pm \omega E_3}$.  The energy eigenvectors were given in a previous paper \cite{KRR1}.  They are not eigenvectors of $\Sigma_3$ unless the momentum is parallel or anti-parallel to the $z$ axis.

What effect does vorticity have on the equilibration time?  Let $\tau_a(E_a,\omega)$ denote the equilibration time with vorticity $\omega$.  Neglect Bose enhancement and Pauli suppression in the final state in Eq. (\ref{eq:RTA:relaxationtime}), and let $a$ be a strange quark.  If it scatters from a gluon $b$ there is no effect on $\tau_a$ due to vorticity.  If it scatters from an up or down quark or antiquark $b$ with equilibrium distribution function approximated by 
$\exp(-E_{\pm}/T)$ there is an extra factor behind the integration sign in Eq. (\ref{eq:RTA:relaxationtime}) of
\bd
\cosh\left( \frac{\omega E_{b3}}{2 E_b T}\right)
\ed
when $\omega \ll E_b$.  With the numbers given above this gives a 0.02\% correction at most and totally ignorable.  There is a similarly insignificant change to the strange quark equilibrium distribution function on the left side of Eq. (\ref{eq:RTA:relaxationtime}).

\section{Conclusion}

In our quest to understand the polarization of $\Lambda$ and $\overline{\Lambda}$ in non-central heavy ion collisions we have studied various mechanisms which could cause strange quark spin equilibration by coupling it to the fluid vorticity \cite{KRR1,KRR2}.  Based on the tetrad formalism we derived, from first principles, the Lagrangian for a Dirac particle in the rest frame of the fluid under rotation. This Lagrangian does not contain a quark-gluon vertex that changes the spin or helicity. Thus, there are no new mechanisms apart from the ones we have already considered. In addition, we compared the time scales for spin versus helicity equilibration and found them to be the same provided fluid vorticity is small compared to the temperature, the typical scenario in heavy ions collisions.  It is readily checked that the introduction of quark chemical potentials does not change the answers to the three questions we posed in this paper.

\section*{Acknowledgement}
The work of JIK was supported by the U.S. Department of Energy Grant DE-FG02-87ER40328.  The work of ER was supported by the U.S. National Science Foundation Grant PHY-1630782 and by the Heising-Simons Foundation Grant 2017-228.


\end{document}